\documentstyle[editedbook,epsfig,psfig,epsf]{mq}
\def\eg{{\it e.g.} }

\def\cm2{cm$^2$ }
\def\se1{s$^{-1}$ }

\begin{opening}
\title{Post-outburst radio spectral evolution of Cygnus X-3}
\author{James C.A. Miller-Jones$^1$, Katherine M. Blundell$^{1}$,
Michael P. Rupen$^2$,}
\author{Anthony J. Beasley$^3$ \& Amy
J. Mioduszewski$^2$} \institute{$^1$ Oxford University Astrophysics,
Keble Road, Oxford, OX1 3RH, UK.\\ $^2$ NRAO, Array Operations Centre,
1003 Lopezville Road, Socorro, NM 87801, USA.\\ $^3$ Owens Valley
Radio Observatory, Caltech, Big Pine, CA 93513, USA.}
\end{opening}

\runningtitle{Workshop Proceedings}
\runningauthor{Miller-Jones et al}

\begin{document}
\vspace{-0.5cm}
\begin{abstract}
{\small Multifrequency VLA and OVRO observations of the radio outburst
of Cygnus X-3 in September 2001 are presented, illustrating the
evolution of the spectrum of the source over a period of six days.
An estimate of the magnetic field in the emitting region is made from
the spectral turnover and possible explanations for the spectral
evolution are suggested.}
\end{abstract}

\section{Introduction}

Cygnus X-3 is an X-ray binary system located in the Galactic plane at
a distance of $\approx$10\,kpc (\eg \cite{Pre00}, \cite{Dic83}) in or
behind one of the spiral arms.  There is ongoing debate as to the
nature of the compact object (\eg \cite{Mit98},\cite{Sch96}).  The
large interstellar extinction to the source precludes any optical
spectroscopy, making it difficult to obtain a reliable mass function
and hindering identification of the companion star. Analysis of
infrared spectra \cite{vanKer96} suggests that the companion is a WN7
Wolf-Rayet star, although more recent observations \cite{Fuc02}
indicate a WN8 subclass.  The orbital period, as confirmed in X-ray
and infrared flux modulations, is 4.8 hours (\eg \cite{Par72}).  The
source occasionally undergoes huge radio outbursts where the flux
density increases to a level of up to 20\,Jy.  Two-sided jets have
been seen on arcsecond scales in a N-S orientation \cite{Mar01},
whereas a highly-relativistic ($\beta \geq 0.81$) one-sided jet with
the same orientation has been reported on milliarcsecond scales with
the VLBA \cite{Mio01}.

\section{Observations of spectral evolution}
Cygnus X-3 went into flare during September 2001, with the RATAN
monitoring programme detecting the start of the outburst on 14th
September \cite{Tru02}.  Brief 2-3 minute VLA snapshot observations
were made every day between 18th and 24th September, and then every
few days until the end of October.  The VLA was in its most compact
``D'' configuration, meaning that the source was unresolved, yet that
none of the emission was resolved out.  Data were taken at seven
different frequencies between 74\,MHz and 43\,GHz.  Between the 20th and
26th September, data were also taken at 100\,GHz at OVRO, extending the
frequency coverage to span three decades in frequency.  In addition,
six longer observations were made with the VLBA between the 18th and
the 23rd September, and appear as before to show a one-sided jet
\cite{Mio02}.

\begin{figure}[h]
\vskip 1cm
\includegraphics{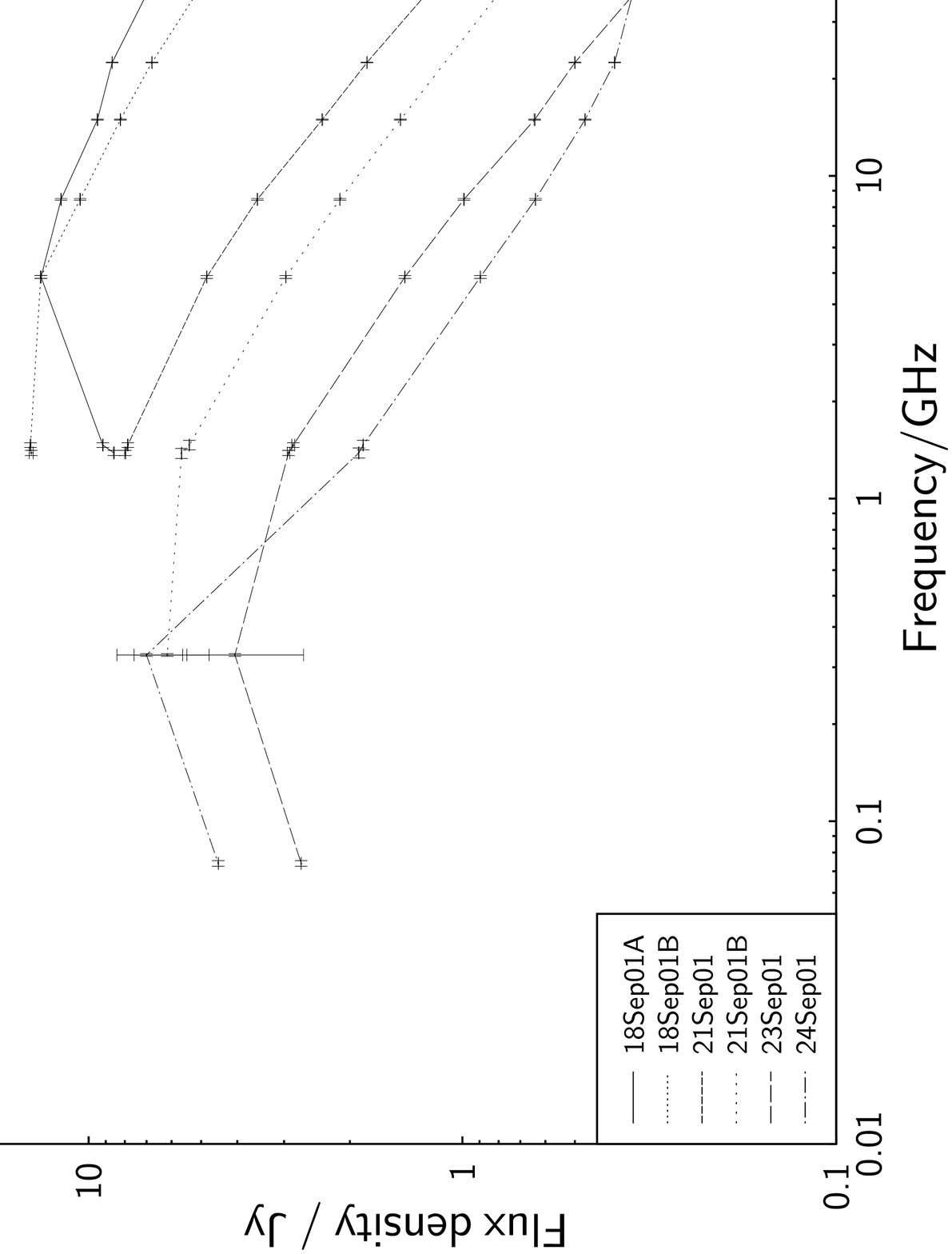}
\vskip 5.5cm
\begin{verse}
\caption{Spectra of Cygnus X-3 during the first week of observations following the flare of September 2001}
\end{verse}
\label{fig:spectra}
\end{figure}

The high-frequency spectral indices taken in the first week of
observations are shown in Figure \ref{fig:spectra}.  The spectra
display the power-law form characteristic of synchrotron emission at
high frequencies, together with a low-frequency cutoff, which moves to
lower frequencies as time progresses.

\begin{figure}[h]
\vskip -1cm
\includegraphics{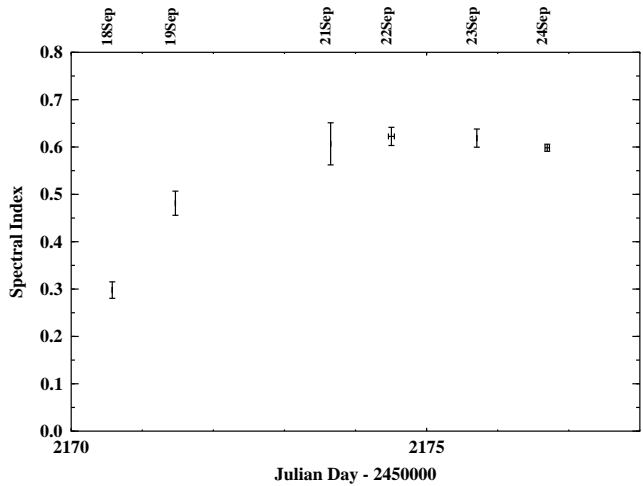}
\vskip 7.5cm
\begin{verse}
\caption{Evolution of the spectral index of Cygnus X-3 during the flare of September 2001}
\end{verse}
\label{fig:specindex}
\end{figure}

After the outburst the source flux decreases, and the spectral
index of the power law increases from $\alpha=0.3$ on the 18th to a
terminal value of $\alpha=0.6$ (implying an electron index of $p=2.2$)
by the 21st September, after which $\alpha$ remains approximately
constant.  This is shown in Figure \ref{fig:specindex}.  Such
behaviour (the increase during the outburst of the spectral index to a
terminal value) was observed in the outburst of 1972 (\cite{All72},
\cite{Hje72}, \cite{Pet73}, \cite{Sea74}).  Various explanations were
invoked, including synchrotron losses \cite{Mar75}, repeated sporadic
electron injection \cite{Hje72,Pet73}, and bremsstrahlung losses
\cite{All72}.  It has also been seen in later outbursts (\eg
\cite{Gel83}, \cite{Tru02}).

The steepening of the spectral index cannot be explained by adiabatic
expansion of the source (\eg \cite{vanderLaa66}, \cite{Sch68}), and is
usually explained in extragalactic sources by invoking synchrotron losses.
However, in this case, the radiative lifetime of the electrons is far
too long (of order $10^8$ years) given the magnetic field estimated in
section \ref{sec:srcparms}.  Moreover, the standard synchrotron loss
models all produce a break in the spectrum \cite{Car91}.  Since the
spectrum is sampled over three decades in frequency, and appears to
maintain a power law, this explanation must be rejected.

The standard explanation for a flat-spectrum source is that it is
composed of a superposition of self-absorbed components
(\cite{Cot80}).  While this can explain a single flat spectrum, it
would seem difficult to ensure that the components evolve in such a
way as to both preserve a power-law form, and steepen the spectral
index gradually.  Initial attempts to model this scenario indicate
that the initial conditions would need to be very contrived in order
to achieve this.

Particle injection would steepen the spectrum, but again on a
timescale of $10^8$ years, as synchrotron losses take effect (as in
the continuous injection model \cite{Car91}).  However, the injected
particles would radiate, so unless the injection parameters varied
with each injection event in a contrived manner, the same difficulties
arise as for the superposition model.

One possibility is that bremsstrahlung emission from the
synchrotron-emitting population is causing Coulomb cooling of the
electrons.  Since the power emitted in bremsstrahlung radiation scales
as $\rho^2$, where $\rho$ is the ionised plasma density, then as the
source expands, $\rho$ decreases and the effect becomes less
significant.  This could also in principle explain the cessation of
the steepening of the spectral index.  The energy loss rate by
bremsstrahlung emission for ultra-relativistic electrons in a fully
ionised plasma is proportional to the electron energy.  This would
ensure that the highest-energy electrons lost energy fastest, leading
to a steepening of the spectral index as required.  A similar scenario
was first put forward by \cite{All72}, who also invoked neighbouring
thermal plasma to generate the bremsstrahlung emission.  While this is
plausible, given the strong stellar wind of the Wolf-Rayet companion,
electron motion between mirror points in an extended stellar
atmosphere is additionally required in their model.

We are currently exploring other physical models which replicate the
details of Cygnus X-3's behaviour in intensity and spectral evolution
over the appropriate timescales.  

\section{Source parameters}\label{sec:srcparms}
As time passes, the low-frequency turnover gradually moves to lower
frequencies.  Assuming that this is due to synchrotron
self-absorption, and fitting the data with a steep and an inverted
power-law, the turnover frequency may be found, given the flux density
at this frequency.  From \cite{Ghi92}, these parameters, together with
a source size, may be used to place an upper limit on the magnetic
field in the plasmon.
\begin{equation}
B=10^{-5}b(\alpha)\theta_{\rm d}^4\nu_{\rm m}^5F_{\rm m}^{-2}\delta_{\textrm{max}}
\textrm{G}
\end{equation}
where $\theta_{\rm d}$ is the angular diameter of the source in mas, $\nu_{\rm m}$
is the turnover frequency in GHz, and $F_{\rm m}$ is the source flux density
(in Jy) at the turnover frequency.  $b(\alpha)$ is a dimensionless
function tabulated in \cite{Ghi92}, $\delta_{\textrm{max}}$ is a
beaming parameter, taken as 2 \cite{Rey96}, and the angular size is
assumed to be 1\,mas \cite{Mio01}.  With the fitted turnover
frequencies and flux densities, the turnover was found to be at
2.4\,GHz on the 18th and at 0.74\,GHz on the 23rd, giving a magnetic
field of 12\,$\mu$G on the 18th and 5\,$\mu$G on the 23rd.  Assuming
flux conservation holds, $B \propto r^{-2}$, which implies expansion
of the plasmon by a factor 1.6 over the 5 days.  As a consistency
check on the magnetic field, it is possible to use the ratios of the
flux densities and the frequencies of the self-absorption turnovers on
the two days to derive an expansion factor $F$ \cite{Sch68}.
\begin{equation}
\Delta(\log\nu) = 2\log F
\end{equation}
\begin{equation}
\Delta(\log S_{\nu}) = 4\log F
\end{equation}
This method gives an expansion factor $F$ of 1.3 and 1.9 respectively
for the frequency and flux estimates.  This is remarkably consistent
with that derived from the magnetic field and inspires confidence in
the field derivation, since the methods are wholly independent.

\section*{Acknowledgements}
JMJ thanks the workshop organisers, St. John's College, Oxford, and
Oxford Astrophysics for support.  This work is supported by a PPARC
studentship.  We also thank the organisers for a very stimulating
workshop.

\end{document}